# COHERENT RADIATION IN INSERTION DEVICES

*A.A. Mikhailichenko, Cornell University, CLASSE, Ithaca, New York, 14853*
*E.G.Bessonov, Lebedev FIAN, Moscow, 119991*

*Abstract.* We calculate the coherent radiation in an undulator/wiggler with a vacuum chamber of arbitrary cross section. The backward radiation is a coherent and it has wavelengths about twice the period of the undulator/wiggler. Mostly of coherent radiation is going with the wavelengths approximately the bunch length at small angles however.

Classification 3: Linear Colliders, Lepton Accelerators and New Acceleration Techniques, T19.
Work supported in part by NSF grant PHY-0202078, RFBR Grant 09-02-00638a

## INTRODUCTION

The period of undulators in newly designed installations (storage rings or FELs) becomes smaller and smaller, while the *inner* vacuum chamber in wigglers sometimes happens to be large or absent at all. Therefore the coherent radiation (CR) might manifest itself here. In some publications, see for example [1]-[3], these conditions investigated to some extent. Calculation of CR in a free space could be based of formulas for radiation of a single particle [4]. In a vacuum chamber of arbitrary cross sections such calculations becomes very laborious, see [2], [5]-[8].

## CR IN A FREE SPACE

The particle oscillating in an undulator with a spatial period $\lambda_u = 2\pi \bar{\lambda}_u$ radiates on harmonics [6], [7]

$$\omega_k = \frac{k\Omega}{1-\vec{n}\vec{\beta}} = \frac{k\Omega}{1-n\beta_\| Cos\theta}, \quad (1)$$

where $k=1,2,...$, $\Omega = \beta_\| c/\bar{\lambda}_u$, $\bar{\beta} = \bar{v}/c \cong \beta_\|$, $\bar{v}$ is a particle's average longitudinal velocity in the undulator. Formula (1) corresponds to the Doppler-shifted frequency $\Omega$. Accordingly, the wavelength of radiation is

$$\bar{\lambda}_k = \frac{\bar{\lambda}_u \cdot (1-\beta_\| Cos\theta)}{k\bar{\beta}} \quad (2)$$

The energy distribution of undulator radiation (UR) emitted by a single particle in an undulator of length $L_u = M\lambda_u$ (M – is the number of periods), during the time duretion $\Delta t = 2\pi M/\Omega$ is defined by the expression [6], [7]

$$\frac{d\varepsilon_k}{do} = \frac{e^2\omega_k^3}{ck\Omega^2}\left[\beta_\perp^2 J_k'^2\left(\frac{k\beta_\perp Sin\theta}{1-\bar{\beta}Cos\theta}\right) + \frac{(Cos\theta-\bar{\beta})^2}{Sin^2\theta}J_k^2\left(\frac{k\beta_\perp Sin\theta}{1-\bar{\beta}Cos\theta}\right)\right] \quad (3)$$

where $\beta_\perp = v_\perp/c = K/\gamma$, $v_\perp$ is the transverse velocity. In the dipole approximation, K<1, $\beta_\| = \sqrt{\beta^2 - \beta_\perp^2} \cong \beta$ and in the ultra-relativistic case $\gamma >> 1$, only the first harmonic radiated, k=1.

$$\frac{d\varepsilon_1}{do} = \frac{e^2\Omega M}{c}\cdot F(\theta) \cong \frac{e^2\Omega M}{c}\cdot \frac{\beta_\perp^2}{8\pi(1-\bar{\beta}Cos\theta)^3}\left[1+\frac{(Cos\theta-\bar{\beta})^2}{(1-\bar{\beta}Cos\theta)^2}\right] \quad (4)$$

where the function $F(\theta)$ introduced accordingly. The energy, radiated by a bunch with population $N_b$ within angles $\{\pi, \theta_m\}$ can estimated by integrating (4) over the solid angle

$$\Delta\varepsilon_1 = N_e \int_{\theta_m}^{\pi} \frac{d\varepsilon_1(\theta)}{do}do = 2\pi N_e \int_{\theta_m}^{\pi} \frac{d\varepsilon_1(\theta)}{do}Sin\theta d\theta \quad (5)$$

In case $\theta_m = 0$, formula (5) gives the total radiated energy at the first harmonic in dipole approximation

$$\Delta\varepsilon_{1tot} = N_e \tfrac{2}{3}r_e^2\overline{B^2}\gamma^2 M\lambda_u = \tfrac{8\pi^2}{3}e^2 N_e M K^2\gamma^2/\lambda_u \quad (6)$$

For the energy radiated coherently, one should suggest $\theta_m = \theta_{coh}$, where the angle of coherence $\theta_{coh}$ is defined from (2), [8]. By suggesting that the wavelength of radiation coincides with the bunch sigma $\sigma_b$,

$$\sigma_b \cong \bar{\lambda}_{coh} = \bar{\lambda}_u \cdot (1-\beta_\| Cos\theta_{coh})/\bar{\beta} \quad (6)$$

one can find that the energy radiated coherently, becomes

$$\Delta\varepsilon_{1coh} = N_e^2 \int_{\theta_{coh}}^{\pi} \frac{d\varepsilon_1(\theta)}{do}do = \frac{2\pi e^2 N_e^2 M\Omega}{c}\int_{\theta_{coh}}^{\pi} F(\theta)Sin\theta d\theta \cong$$

$$\cong \frac{e^2 N_e^2 M\Omega\beta_\perp^2}{4c\bar{\beta}}\frac{1}{(1-\bar{\beta}Cos\theta)^2}\Big|_\pi^{\theta_{coh}} \cong \frac{e^2 N_e^2 M\Omega\beta_\perp^2 \bar{\lambda}_u^2}{4c\bar{\beta}\sigma_b^2} \quad (7)$$

Within the approximation accepted above

$$\frac{\Delta\varepsilon_{1coh}}{\Delta\varepsilon_{1tot}} \cong \tfrac{3}{4}\frac{N_e}{\gamma^4}\left(\frac{\lambda_u}{\sigma_b}\right)^2 \quad (8)$$

Generally the energy-loss ratio is

$$\frac{\Delta\varepsilon_{1coh}}{\Delta\varepsilon_{1tot}} \cong N_e \int_{\theta_{coh}}^{\pi} F(\theta)do \Big/ \int_0^{\pi} F(\theta)do, \quad (9)$$

and could be evaluated with MATHEMATICA. Some results are represented in Fig.1. We would like to underline here that for a single electron the ratio of the total energy radiated in a forward and in a backward direction is $\gamma^4$ in full agreement with formula (73,11) from [4].

*Estimations*

For the energy of $5 GeV$ ($\gamma \cong 10^4$), and for the bunch sigma $\sigma_z \cong \lambda_{coh} = 10^{-2}cm$, $\lambda_u = 2cm$, according to (8), the ratio

$$\Delta\varepsilon_1 / \Delta\varepsilon_{coh} \cong 10^{12}/N_e \quad (10)$$

So the losses of the coherent radiation become equal to the incoherent forward radiation for the bunch population

$N_e \approx 10^{12}$, i.e. for the bunch charge $Q_b \cong 160\ nC$. In that case, the loss per particle becomes twice of what it would be for a single particle.

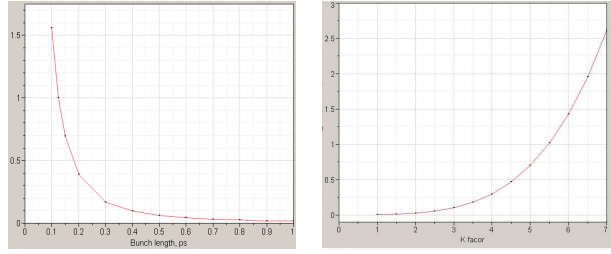

Figure 1. Ratio of the coherent/incoherent energy loss as a function of the bunch length, *ps*; *K*=2, *Q*=100*pQ*, left. At the right: the ratio of the coherent/incoherent energy loss as a function of *K*, $\sigma_b/c = 2\ ps$.

A more accurate evaluation of the coherently radiated energy with formulas (2)-(6) as a portion of the energy radiated within angles $\{\pi \div \theta_{coh}\}$ multiplied by $N_e$ versus total incoherent radiation (within angles $\{\pi \div 0\}$) as a function of the bunch length or *K* factor is shown in Fig.1.

It is interesting, that for the beam duration of approximately 0.1*ps*, *Q*=100*pC*, the ratio of coherent to incoherent energy losses comes to ~0.6 for an undulator with *K*=1.5, $\lambda_u = 2cm$ at 5 *GeV*; for *K*=3 the coherent loss is 6.5 times bigger, than the incoherent one. For 1 *nQ* this ratio comes to 65 for the last set of parameters. For *atto*second bunches radiation is purely coherent and it has the wavelengths on the order of the bunch length, which makes usage of such bunches problematic for radiation with desired properties. All these were preliminary estimations. For a more correct evaluation of the effect in a chamber of arbitrary cross section we used the FlexPDE solver.

## CR IN A VACUUM CHAMBER

In the presence of vacuum chamber with the cross section *a*x*a*, the wavelength of radiation $\lambda_{wg}$ becomes longer, than the wavelength $\lambda_0$, measured in free space, $\lambda_{wg} = \lambda_0 / \sqrt{1 - (\lambda_0/\lambda_{cr})^2}$, where $\lambda_{cr} \cong 2a$. Radiation can propagate within tan angle that is not smaller than the one, defined by the critical wavelength in a waveguide $\theta_{cr} \geq arcCos(\lambda_0/\lambda_{wg})$ [9], [10]. On the other hand, according to (2), the angle corresponding to the coherence length $\lambda_{coh} \cong \sigma_b$ is $\theta_{coh} \geq arcCos\ (1 - \bar{\beta}\sigma_b/\lambdabar_u)$. This should be larger, than $\theta_{cr}$.

In a chamber with arbitrary cross section, the allowed wavelengths have rather complicated structure, so analytical solution can only be obtained for the simplest cases, such as vacuum chambers of either round or rectangular cross section. We therefore developed a code able to evaluate radiation in a conducting chamber of arbitrary cross section.

The current density was represented as the following
$$\vec{j} = \rho \cdot \vec{v} \cong \rho \cdot (v_x, v_y, v_z), \quad (11)$$
where the charge density
$$\rho = \frac{Q_b}{\pi^{3/2}\sigma_x\sigma_y\sigma_b}\exp\left(-\frac{(z+z_0-\beta ct)^2}{\sigma_b^2}\right)\exp\left(-\frac{(x-x_0)^2}{\sigma_x^2}\right)\exp\left(-\frac{y^2}{\sigma_y^2}\right), \quad (12)$$
and $v_x = cK/\gamma \cdot Cos(\beta ct/\lambdabar_u)$, $v_y = cK/\gamma \cdot Sin(z/\lambdabar_u)$ or $v_y = 0$ for the planar undulator; $v_z \cong c \cdot (1 - \frac{1}{2}K^2/\gamma^2)$, $\sigma_x, \sigma_y$ are the transverse sigmas in *x* and *y* direction. The current distribution (11), (12) allows modeling the coherent part of radiation only. We solved equations for the vector and scalar potentials linked by the Lorentz gauge condition
$$\Delta\vec{A} - \frac{1}{c^2}\ddot{\vec{A}} - \delta \cdot \dot{\vec{A}} = -\mu_0\vec{j}, \quad \Delta U - \frac{1}{c^2}\ddot{U} = -\frac{1}{\varepsilon_0}\rho, \quad \vec{\nabla}A = \frac{1}{c^2}\dot{U} \quad (13)$$
with the boundary conditions $U|_\Sigma = 0$ for the scalar potential; for the magnetic vector potential, tangential component was chosen to be zero on the boundary, similarly for its normal derivative. This coincides with the requirements derived in [11]. The mesh propagates in accordance with the bunch position. The term with decrement, $\delta \cdot \dot{\vec{A}}$, $\delta = 1/c^2\tau$, with $\tau \sim 1ns$, is introduced to describe the losses in the walls. Although the losses occur on the surface, the volume losses introduced by such a way are self consistent. In our model of vacuum chamber, the end section, at ~5% along the *z*-distance, has increased losses, ~100$\delta$ to avoid reflections from the entrance boundary (this section marked by the blue color in Fig. 2).

Chambers with different shapes and dimensions were investigated, including the simplest one with rectangular cross section. The wavelength of the undulator field $\lambda_u$ was varied from 1 to 3 cm. The bunch length was varied also.

Design of vacuum chamber for wigglers is sometimes problematic, as it contains so-called cleaning electrodes, what makes this chamber *multiply* connected [10]. As the backward radiation is 100% coherent and goes at central wavelength equal to the doubled period of wiggler, one may expect strong excitation of undesirable fields there.

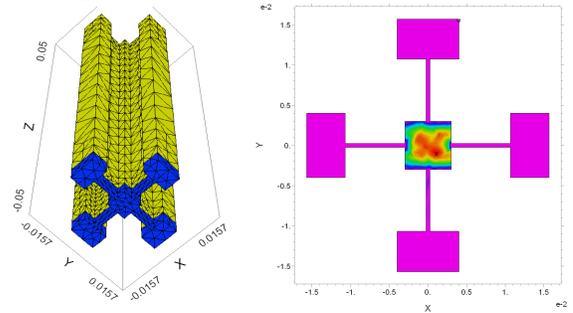

Figure 2. 3D mesh for the vacuum chamber used in modeling. At the right: The modulus of electric field for the chamber at one particular moment (frame from a movie).

In Figure 3 the bunch length is 2.5*mm*, the period of the undulator is 15 mm. Here it is clearly seen that radiation propagates in side slits (see Fig.2).

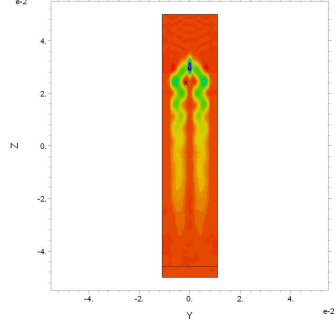

Figure 3. Contour plot for the vector-potential. The bunch is moving in a chamber from Fig.2, with sub-critical transverse dimensions of the central part (2x2*mm*$^2$); (single frame from a movie).

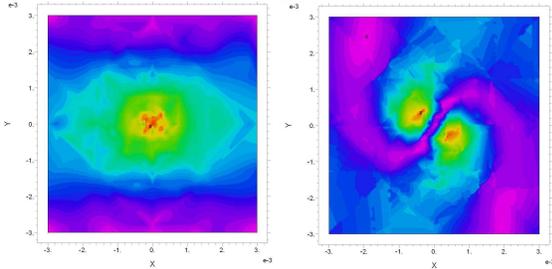

Figure 4. Electrical (at the left) and magnetic (at the right) fields modulus in a rectangular chamber of 6x6 *mm*$^2$ cross-section of a helical undulator with left helicity.

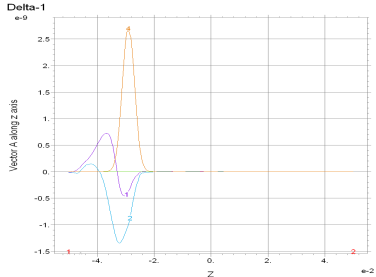

Figure 5. Components of the vector potential along *z*. The bunch is moving from the left to the right (frame from a movie)

Energy lost by the bunch evaluated by taking the integral

$$\Delta \varepsilon = \int dt \int (\vec{j} \cdot \vec{E}) dV = -\int dt \int (\vec{j} \cdot \frac{\partial \vec{A}}{\partial t}) dV, \qquad (14)$$

where only the transverse components of vector-potential (and $\vec{E}$) are taken into account; this corresponds to pure radiation field.

## CONCLUSION

Coherent radiation in the latest wigglers/undulators might represent a problem if designed without its consideration. For planned bunches with femtosecond durations, practically all radiation will be concentrated in sub-millimeter wavelengths. For x-ray ERLs, the coherent radiation is in transition regime and requires attention for some modes of operation. It was found that wide slits between the magnets (see [12]) support waves with wavelengths larger, than the critical ones for the central core, as it was expected. In addition the CSR is sensitive to the exact dimensions of these slits, as they demonstrate resonant properties. For a bunch with charge 100*pC*, time duration of100*fs*, the ratio of the energy lost by coherent radiation to the energy lost incoherently may reach approximately 2.2 for the undulator with $\lambda_u = 2.4 cm$, *K*=2. For the central core dimensions less, than the critical ones, the radiation propagates exclusively in the slits. In this case the transverse size of the beam becomes dominating parameter. This is in agreement with the physical expectations, as in this case the radiation is emitted to the sides ($\theta \approx \pi / 2$), so that transverse dimensions define the coherent threshold.